\documentclass{blois}

\def\be{\begin{equation}}
\def\ee{\end{equation}}
\def\bea{\begin{eqnarray}}
\def\eea{\end{eqnarray}}

\newcommand{\Photo}{\includegraphics[height=35mm]{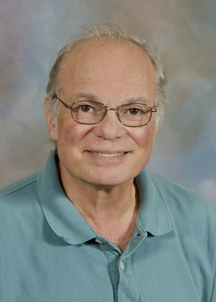}}

\begin{document}
\vspace*{4cm}\title {EXTRACTION OF $\sin^2\theta^{\rm lept}_{\rm eff}$  AND INDIRECT MEASUREMENT OF $M_W$ 
FROM  THE 9 FB$^{-1}$ FULL RUN II SAMPLE OF  $\mu^+\mu^-$ EVENTS AT CDF}

\author{ A. Bodek}

\address{for the CDF collaboration\\ Department of Physics and Astronomy, University of Rochester, Rochester, NY. 14627, USA\\
Presented at 26th Rencontres de Blois, Particle Physics and Cosmology,  May  18-23, 2014}

\maketitle\abstracts{
We report on the extraction of $\sin^2\theta^{\rm lept}_{\rm eff}$ and indirect measurement of
the mass of the W boson from the forward-backward asymmetry of $\mu^+\mu^-$
events in the $Z$ boson  mass region.  The data sample collected by the CDF detector
corresponds to the full 9 fb$^{-1}$  run II sample.  
We measure   $\sin^2 \theta^{\rm lept}_{\rm eff}  =  0.2315 \pm 0.0010$, 
 $ \sin^2 \theta_W  =  0.2233 \pm 0.0009$ and   
  $M_W ({\rm indirect})  =  80.365 \pm 0.047 \;{\rm GeV}/c^2$,
where each uncertainty includes both statistical and systematic
contributions. }

\section{Introduction}
\begin{figure}
\centerline{\includegraphics[height=60mm,width=1.\linewidth]{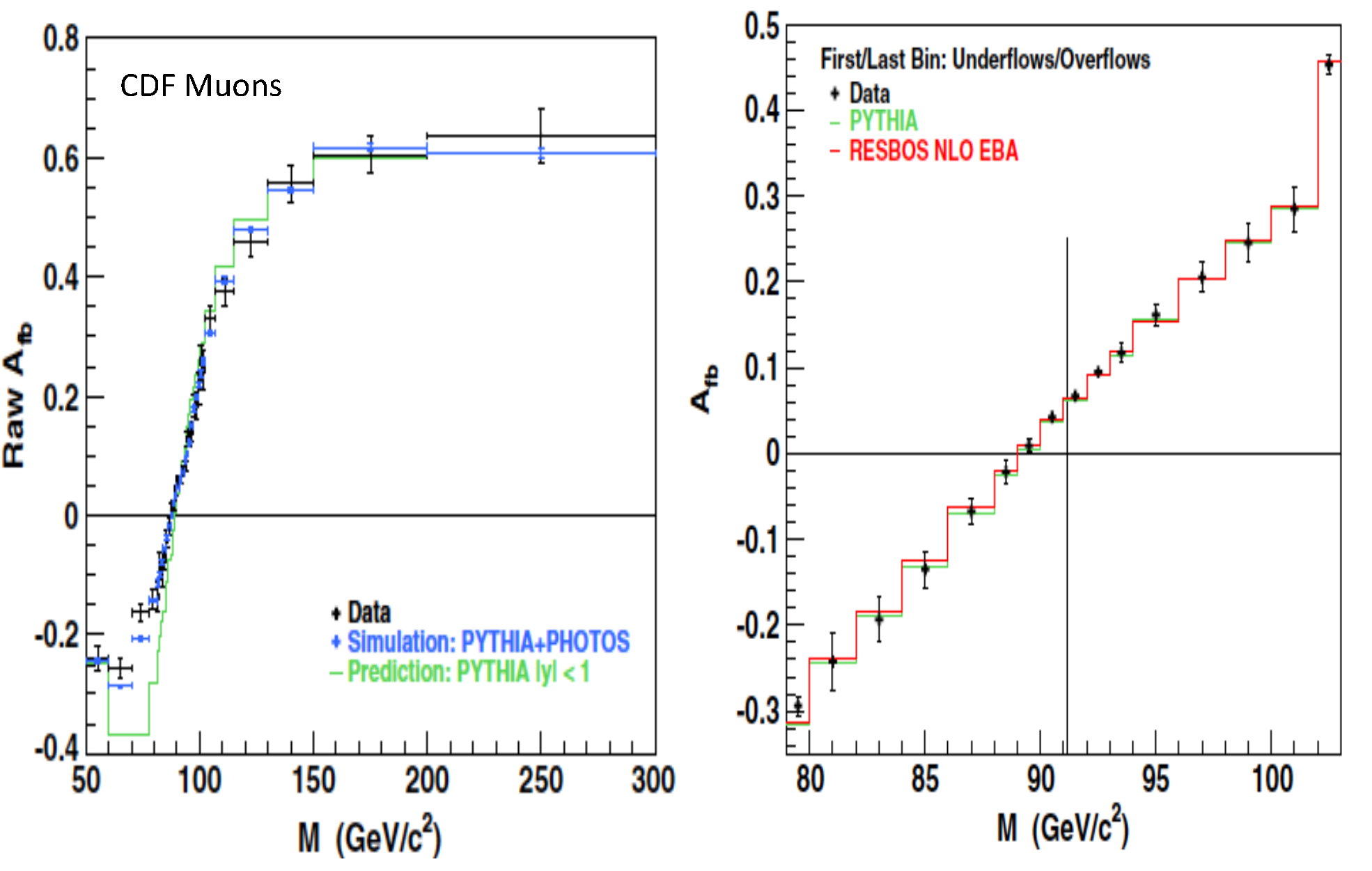}}
\centerline{\includegraphics[height=62mm,width=1.\linewidth]{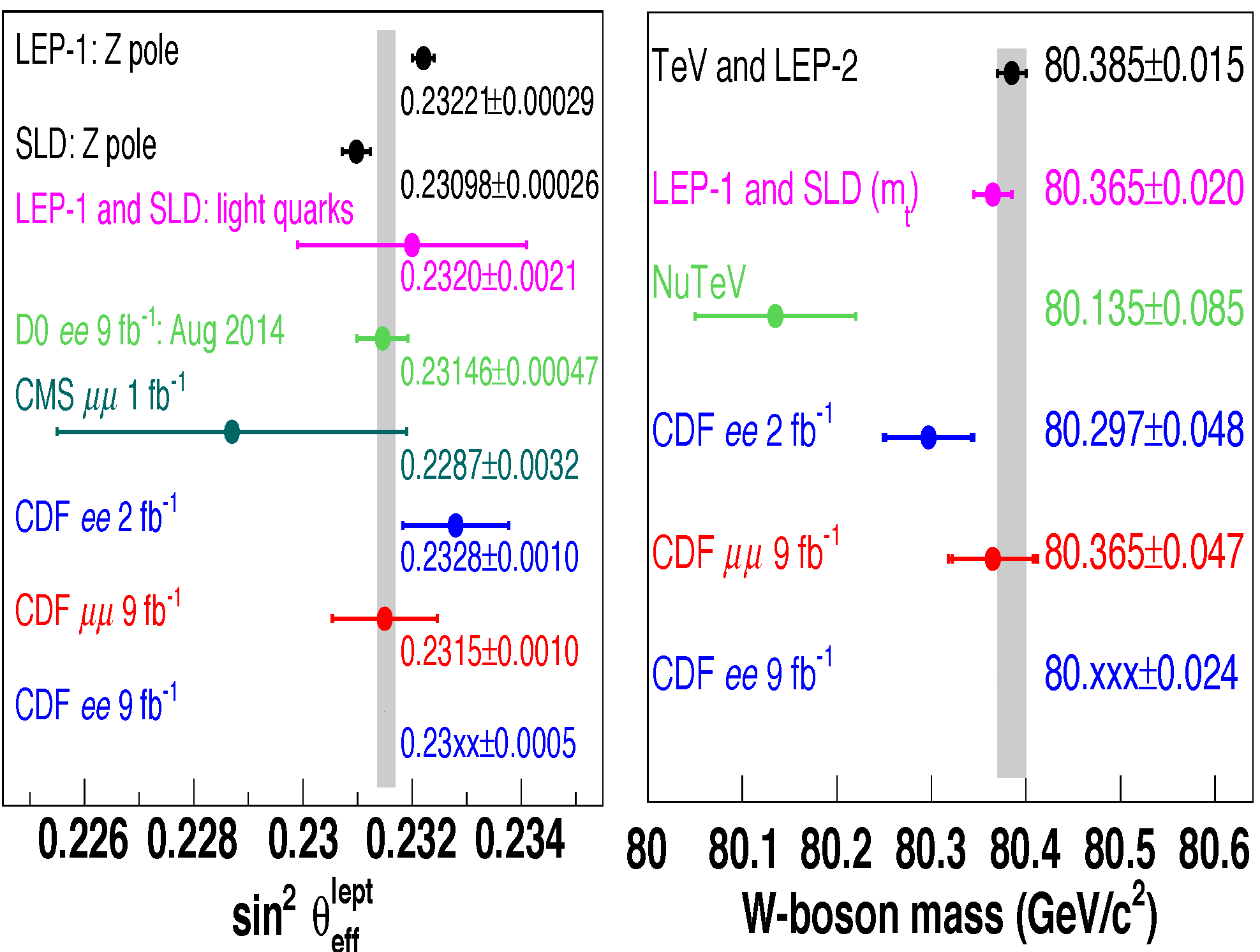}}
\caption{ {\bf Top-left}: CDF raw $A_{\rm fb}$ measurement in bins of $\mu^+\mu^-$
invariant mass.  Only statistical uncertainties are shown.
The Monte Carlo
simulation (\textsc{pythia}) includes the effect of resolution smearing and FSR.
The \textsc{pythia} $|y|<1$ asymmetry curve does not.
 {\bf Top-right}:  $A_{\rm fb}$ unfolded for resolution and QED-FSR.  The \textsc{pythia} calculation uses
$\sin^2\theta^{\rm lept}_{\rm eff} = 0.232$.
The EBA-based \textsc{resbos} calculation uses
$\sin^2\theta_W = 0.2233$ 
$(\sin^2\theta^{\rm lept}_{\rm eff} = 0.2315)$.$~~~~~$
 {\bf Bottom-left}:  Measurements of $\sin^2\theta^{\rm lept}_{\rm eff}$. {\bf Bottom-right}: Direct and indirect
measurement of  $M_W$.   Also shown are the errors from  9 fb$^{-1}$ $e^+e^-$ sample in CDF  which are
expected to be smaller by a factor of 2.
}
\label{fig-comparison}
\end{figure}

Now that the Higgs mass is known, the Standard Model is over constrained. Therefore, any inconsistency between precise measurements  of SM parameters would be indicative of new physics.  The parameter that needs to be measured more precisely is  $M_W$ (with errors $<$15  MeV), or equivalently 
$\sin^2\theta_W= 1 - M_W^2/M_Z^2$ (with errors  $<$0.0003).  Similarly, in order to help resolve the  long standing  $3\sigma$ difference in   
 $\sin^2\theta^{\rm lept}_{\rm eff}$   between SLD and LEP (shown in bottom left part of  Figure~\ref{fig-comparison}), new measurements of   $\sin^2\theta^{\rm lept}_{\rm eff}$ should have errors similar to SLD or  LEP ($\pm$0.0003).
Precise extractions of $\sin^2\theta^{\rm lept}_{\rm eff}$and  $\sin^2\theta_W= 1 - M_W^2/M_Z^2$  using the
forward-backward asymmetry  ($A_{\rm fb}$) of dilepton events produced in 
p$\bar p$ and pp collisions  are now  possible for the first time because of  three new innovations:
\begin{itemize}
\item The  use a new technique \cite{muon-scale} for calibrating the muon energy scale
as a function of  detector  $\eta$ and $\phi$ (and sign), thus greatly reducing systematic            
errors from the energy scale.  A similar method can also used
for electrons.

\item The  implementation\cite{cdf-ee} of Z fitter electroweak radiative corrections into the theory predictions of POWHEG and RESBOS 
 which 
 allows for a measurement of both  $\sin^2\theta^{\rm lept}_{\rm eff}$(M$_Z$) and 
  $\sin^2\theta_W= 1 - M_W^2/M_Z^2$.
  
\item Use of a new  event weighting technique\cite{event-weighting}.  With this technique
all  experimental uncertainties in acceptance and efficiencies cancel (by measuring  the $\cos\theta$ coefficient $A_4$  and  using the relation
 $A_{FB}=8A_4/3$). Similarly,  additional weights are included for antiquark dilution, which makes the analysis independent of the acceptance in dilepton rapidity. 
\end{itemize}

\section{Analysis of CDF $\mu^+\mu^-$  full 9 fb$^{-1}$  run II sample }
We report on the  published analysis
of the  full 9 fb$^{-1}$  run II $\mu^+\mu^-$  data sample \cite{cdf-mumu} collected by the CDF detector.\cite{cdf-mumu}

After applying the calibrations and muon scale corrections to the experimental
and simulated data, $A_{\rm fb}$ is measured in bins
of $\mu^+\mu^-$ invariant mass using the event-weighting method.
This measurement is denoted as the raw $A_{\rm fb}$
measurement because the event-weighting method provides a first-order
acceptance correction, but does not include resolution unfolding and
final-state (FSR) QED radiation.
The raw $A_{\rm fb}$ measurement in bins of the muon-pair
invariant mass  is shown on the top left part of  Fig.~\ref{fig-comparison}.  Only statistical uncertainties are shown.
The Monte Carlo
simulation (\textsc{pythia}) includes the effect of resolution smearing and FSR.
The \textsc{pythia} $|y|<1$ asymmetry curve does not include
the effect of resolution smearing or  FSR.

With  the event weighting technique, the events near  $\cos\theta$=0 are assigned zero weight, Therefore,
the migration of events between positive and negative $\cos\theta$ is negligible. Resolution smearing
and FSR primarily transfer events between bins in invariant mass.

The raw $A_{\rm fb}$ in bins of invariant is unfolded\cite{cdf-mumu} for resolution smearing and FSR using a transfer matrix
which is obtained from the Monte Carlo simulation. The unfolded $A_{\rm fb}$ is shown in the top-right side of Fig. \ref{fig-comparison}.

The electroweak (EWK) mixing parameters $\sin^2\theta^{\rm lept}_{\rm eff}$ and
$\sin^2\theta_W$ are extracted from the 
 fully unfolded 
$A_{\rm fb}$ measurements using
$A_{\rm fb}$ templates calculated with different values of
$\sin^2\theta_W$. Three  QCD 
calculations are used: LO (tree), \textsc{resbos} NLO, and
\textsc{powheg-box} NLO.  The calculations were modified
to include EWK radiative correction using the
effective Born approximation (EBA.\cite{cdf-ee}
For the EBA electroweak form-factor
calculations, the EW parameter is $\sin^2\theta_W$.
\par
The $A_{\rm fb}$ measurement is directly sensitive to the
effective-mixing parameters $\sin^2\theta^{\rm lept}_{\rm eff}$ which
are combinations of the form-factors and $\sin^2\theta_W$.  Most of
the sensitivity to  $\sin^2\theta^{\rm lept}_{\rm eff}$ comes from the Drell-Yan $A_{\rm fb}$
near the Z pole, where  $A_{\rm fb}$ is small.  In contrast,  $A_{\rm fb}$ at higher mass values  where  $A_{\rm fb}$
is large, is mostly sensitive to the axial coupling, which is known.
\par
\begin{figure}
\centerline{\includegraphics[height=58mm,width=1.\linewidth]{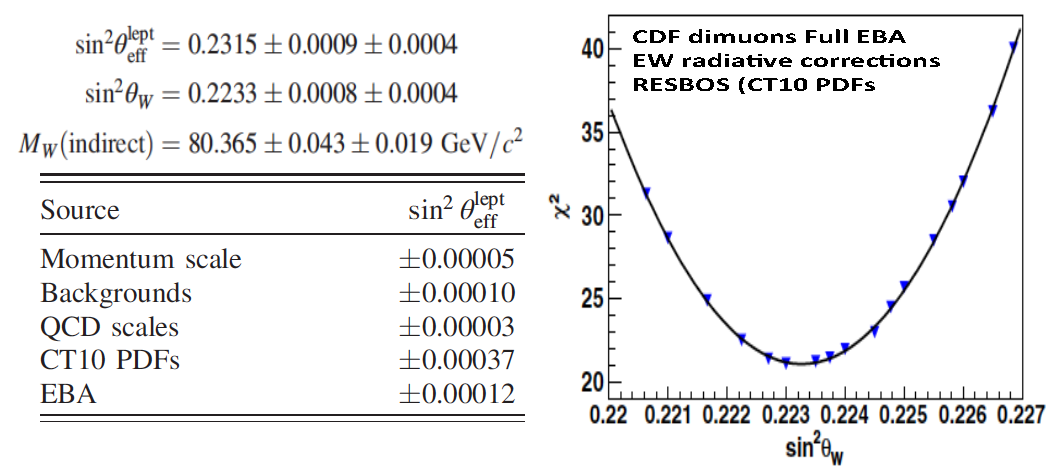}}
\caption{ {\bf Left}: Extracted values of  $\sin^2\theta^{\rm lept}_{\rm eff}$, $\sin^2\theta_{W}$, and $M_W$ from the
CDF measurement of  $A_{\rm fb}$ in the 9 fb$^{-1}$ $\mu^+\mu-$ sample (with sources of systematic errors).
 {\bf Right}:  $\chi^2$ comparison of the CDF $A_{\rm fb}$ $\mu^+\mu-$ measurement with \textsc{resbos}-EBA
NLO templates.   
}
\label{cdf-results}
\end{figure}

The measurement and templates are compared using the $\chi^2$
statistic evaluated with the $A_{\rm fb}$ measurement
error matrix. Each template provides a scan point for the $\chi^2$ function
$(\sin^2\theta_W, \chi^2( \sin^2\theta_W))$. The scan points
are fit to a parabolic $\chi^2$ functional form.
The $\chi^2$ distribution of the scan over templates from the
\textsc{resbos} NLO calculation (with CT10 PDFs) is shown in the right side of 
Fig.~\ref{cdf-results}.
The EBA-based \textsc{resbos} calculations of $A_{\rm fb}$
are used to extract the central value of $\sin^2\theta_W$. The other
calculations are used to estimate the systematic error from
the electroweak radiative corrections and QCD NLO radiation.
\section {Systematic errors  in the extraction of   $\sin^2 \theta^{\rm lept}_{\rm eff}$  from the  full 9 fb$^{-1}$  run II sample}
In all QCD calculations, the mass-factorization and
renormalization scales are set to the muon-pair
invariant mass. To evaluate the effects of different scales,
the running scales are varied independently by a factor
ranging from $0.5$ to $2$ in the calculations.
The largest observed deviation of the best-fit value of
$\sin^2\theta_W$ from the default value is considered to be
the QCD-scale uncertainty. This uncertainty is
$\Delta\sin^2\theta_W({\rm QCD \; scale)} = \pm 0.00003$.
\par
The CT10 PDFs are derived from a global analysis of experimental
data that utilizes 26 fit parameters and the
associated error matrix. In addition to the best global-fit
PDFs, PDFs representing the uncertainty along the eigenvectors
of the error matrix are also derived. For each eigenvector $i$,
a pair of PDFs are derived using 90\% C.L. excursions from the
best-fit parameters along its positive and negative directions.
The difference between the best-fit $\sin^2\theta_W$ values
obtained from the positive (negative) direction excursion PDF
and the global best-fit PDF
is denoted as $\delta^{+(-)}_i$. The 90\% C.L. uncertainty
for $\sin^2\theta_W$ is given by the expression
$\frac{1}{2} \sqrt{ \sum_i (|\delta^+_i|+
                            |\delta^-_i|)^2 }$,
where the sum $i$ runs over the 26 eigenvectors. This value
is scaled down by a factor of 1.645 for the 68.3\% C.L. (one
standard-deviation) uncertainty yielding
$\Delta\sin^2\theta_W({\rm PDF}) = \pm 0.00036$. The PDF error 
is expected to be a factor of 2 smaller with more modern PDFs.
\par
The \textsc{resbos} $A_{\rm fb}$ templates are the default
templates for the extraction of
$\sin^2\theta^{\rm lept}_{\rm eff}$.
The scan with the \textsc{powheg-box} or the tree templates yields
slightly different values for $\sin^2\theta_W$. The difference,
denoted as the EBA uncertainty, is
$\Delta\sin^2\theta_W({\rm EBA}) = \pm 0.00012$.
Although the \textsc{resbos} and \textsc{powheg-box} predictions
are fixed-order NLO QCD calculations at large boson $P_{\rm T}$,
they are all-orders resummation calculations in the low-to-moderate
$P_{\rm T}$ region, which provides most of the total cross
section. The EBA uncertainty is a combination of differences between
the resummation calculations and the derived value of
$\sin^2\theta_W$ with and without QCD radiation.
\par
In summary, the total systematic uncertainties on $\sin^2\theta_W$
from the QCD mass-factorization and renormalization scales, and from
the CT10 PDFs is $\pm 0.00036$.
All component uncertainties (shown in the left side of  Fig. \ref{cdf-results}) are combined in quadrature. 
With the inclusion of the EBA uncertainty,
the total systematic uncertainty is $\pm 0.00038$.

\section{Summary of results from the  9 fb$^{-1}$ $\mu^+\mu^-$ sample}
%
The left side of Fig.~\ref{cdf-results} shows the  best fit  extracted 
values of  $\sin^2\theta_{\rm eff}^{\rm lept}$, $\sin^2\theta_{W}$, and $M_W$ from the
CDF measurement of  $A_{\rm fb}$ in the  9 fb$^{-1}$ $\mu^+\mu-$ sample  with statistical errors, and the various sources of systematic errors.
With statistical and systematic errors added in quadrature, the best-fit values are
\begin{eqnarray*}
  \sin^2 \theta^{\rm lept}_{\rm eff} & = & 0.2315 \pm 0.0010 \\
  \sin^2 \theta_W & = & 0.2233 \pm 0.0009 \\   
  M_W ({\rm indirect}) & = & 80.365 \pm 0.047 \;{\rm GeV}/c^2 \, .
\end{eqnarray*}
The results  for   $\sin^2 \theta^{\rm lept}_{\rm eff}$ are consistent with other
measurements at the $Z$-boson pole, as shown on the  bottom  left part of 
Fig, \ref{fig-comparison}.  The results for $M_W$ are consistent with other direct and
indirect measurements of $M_W$ as shown on the bottom-right side of Fig. \ref{fig-comparison}.

Also shown is the most recent (Aug. 2014)   value\cite{dzero}  of   $\sin^2 \theta^{\rm lept}_{\rm eff}$ extracted from the  full 9.7 fb$^{-1}$  run II $e^+e^-$  sample in D0  \cite{dzero} (0.2315 $\pm$ 0.0005).
The results from the  CDF  full 9 fb$^{-1}$  run II $e^+e^-$  sample  are expected by end of
2014.  
Because of the larger angular acceptance for electrons, the expected error
in  $\sin^2 \theta^{\rm lept}_{\rm eff}$  are smller ($\pm$0.0005). 
The expected  error
in the CDF  extracted value of    $M_W^{indirect} $ ($\pm$ 24 MeV) 
 will be competitive with the direct measurements
 of $M_W$.
The uncertainty in the average of   the CDF and D0  legacy  9 fb$^{-1}$  measurement
 of $\sin^2 \theta^{\rm lept}_{\rm eff}$  will be competitive with LEP and SLC.



\end{document}